\PassOptionsToPackage{unicode}{hyperref}
\PassOptionsToPackage{hyphens}{url}
\documentclass[
  12pt,
]{article}
\author{}
\date{\vspace{-2.5em}}

\usepackage{amsmath,amssymb}
\usepackage{lmodern}
\usepackage{iftex}
\ifPDFTeX
  \usepackage[T1]{fontenc}
  \usepackage[utf8]{inputenc}
  \usepackage{textcomp} 
\else 
  \usepackage{unicode-math}
  \defaultfontfeatures{Scale=MatchLowercase}
  \defaultfontfeatures[\rmfamily]{Ligatures=TeX,Scale=1}
\fi
\IfFileExists{upquote.sty}{\usepackage{upquote}}{}
\IfFileExists{microtype.sty}{
  \usepackage[]{microtype}
  \UseMicrotypeSet[protrusion]{basicmath} 
}{}
\makeatletter
\@ifundefined{KOMAClassName}{
  \IfFileExists{parskip.sty}{%
    \usepackage{parskip}
  }{
    \setlength{\parindent}{0pt}
    \setlength{\parskip}{6pt plus 2pt minus 1pt}}
}{
  \KOMAoptions{parskip=half}}
\makeatother
\usepackage{xcolor}
\IfFileExists{xurl.sty}{\usepackage{xurl}}{} 
\IfFileExists{bookmark.sty}{\usepackage{bookmark}}{\usepackage{hyperref}}
\hypersetup{
  hidelinks,
  pdfcreator={LaTeX via pandoc}}
\usepackage[margin=1in]{geometry}
\usepackage{longtable,booktabs,array}
\usepackage{calc} 
\usepackage{etoolbox}
\makeatletter
\patchcmd\longtable{\par}{\if@noskipsec\mbox{}\fi\par}{}{}
\makeatother
\IfFileExists{footnotehyper.sty}{\usepackage{footnotehyper}}{\usepackage{footnote}}
\makesavenoteenv{longtable}
\usepackage{graphicx}
\makeatletter
\def\maxwidth{\ifdim\Gin@nat@width>\linewidth\linewidth\else\Gin@nat@width\fi}
\def\maxheight{\ifdim\Gin@nat@height>\textheight\textheight\else\Gin@nat@height\fi}
\makeatother
\setkeys{Gin}{width=\maxwidth,height=\maxheight,keepaspectratio}
\makeatletter
\def\fps@figure{htbp}
\makeatother
\newlength{\cslhangindent}
\newlength{\csllabelwidth}
\newlength{\cslentryspacingunit} 
\newenvironment{CSLReferences}[2] 
 {
  \setlength{\parindent}{0pt}
  \ifodd #1
  \let\oldpar\par
  \def\par{\hangindent=\cslhangindent\oldpar}
  \fi
  \setlength{\parskip}{#2\cslentryspacingunit}
 }%
 {}
\usepackage{calc}

\ifLuaTeX
  \usepackage{selnolig}  
\fi

\begin{document}

\begin{center}

\LARGE{\bf{Poisson Modeling and Predicting \\ English Premier League Goal Scoring}}

\vspace{3mm}

\large{Quang Nguyen$^{* \dagger}$}

$^{*}$Loyola University Chicago, Chicago, Illinois

$^{\dagger}$Work completed at Wittenberg University, Springfield, Ohio

\end{center}

\vspace{5mm}

\centering \textbf{Abstract}

\raggedright

\begin{quote}
\hspace{5mm} The English Premier League is well-known for being not only
one of the most popular professional sports leagues in the world, but
also one of the toughest competitions to predict. The first purpose of
this research was to verify the consistency between goal scoring in the
English Premier League and the Poisson process; specifically, the
relationships between the number of goals scored in a match and the
Poisson distribution, the time between goals throughout the course of a
season and the exponential distribution, and the time location of goals
during football games and the continuous uniform distribution. We found
that the Poisson process and the three probability distributions
accurately describe Premier League goal scoring. In addition, Poisson
regression was utilized to predict outcomes for a Premier League season,
using different sets of season data and with a large number of
simulations being involved. We examined and compared various soccer
metrics from our simulation results, including an English club's chances
of being the champions, finishing in the top four and bottom three, and
relegation points.
\end{quote}

\emph{Keywords}: English Premier League, sports statistics, Poisson
process, Poisson regression

\hypertarget{introduction}{%
\section{Introduction}\label{introduction}}

Association football, also commonly known as soccer in America, is
undoubtedly the most widely-played sport in the world. Often referred to
as the ``king of sport,'\,' football can be played almost anywhere, from
grass fields to indoor gyms, streets, parks, or beaches, due to the
simplicity in its principal rules and essential equipment. Europe is
known to be the birthplace of modern football (Johnson 2015), and the
European soccer culture is unlike any other. The Old Continent is home
to numerous top-level professional football leagues, and the English
Premier League (EPL) distinguishes itself because of its competition
quality, overall balance, and popularity. Some of the best football
coaches and players in the world come together to compete for the
prestigious Premier League trophy.

\hspace{5mm} The EPL was founded in 1992, and over the last three
decades, we have witnessed numerous memorable matches and countless
outstanding performances by clubs and their players. The EPL is
currently a competition of twenty English football clubs. At the end of
each season, the bottom three teams get relegated to the second-highest
division of English football, in exchange for three promoted teams. A
Premier League season usually takes place from mid-August to mid-May.
Each team gets to play every other team twice, once at home and once on
the road, hence there are a total of thirty-eight fixtures in a season
for each team ({``Premier League Explained''} 2020).

\hspace{5mm} The most important aspect of the game of football is
indisputably scoring goals. Despite the significance of other factors
like ball possessing or disciplined defending, we have to admit that the
main reason we pay to watch soccer is to see the ball being put in the
back of the net. The rule is very simple: in order to win, you must
score more than your opponent. In the Premier League, each match happens
within the span of ninety minutes (plus stoppage time), and the match
consists of two 45-minutes halves. Each team can get one of these three
results after each match: a win, a draw, or a loss. If there is a draw,
the two clubs receive a point apiece, and for non-drawing matches, the
winner is rewarded with three points and the losing team gets punished
with zero points. Thus the club with the most points at the end of the
year will have their hands on the exquisite EPL trophy, and the total
points also determines the fates of teams in the relegation zone
({``Premier League Explained''} 2020). This makes every single match so
critical, as losing one single point could end up costing a team's
chance of winning a title or remaining in the top tier football league
in England.

\hspace{5mm} In this paper, we attempt to use statistical methods to
model and predict goal scoring and match results in the Premier League.
We will first determine whether notable aspects of goal scoring, namely,
the number of goals scored, the time between goals, and time location of
goals in a match, fit the characteristics of a Poisson process. We will
then use Poisson regression to predict what would happen in the 2018-19
EPL season, for instance, which clubs are more likely to win the title
or get relegated, using different subsets of data from prior seasons.
The paper is outlined as follows: We first introduce the data that we
used for our analyses in Section 2. Next, our methodologies are
described in Section 3. We then spend the next two sections, 4 and 5, on
our two main topics of this research - using the Poisson process to
model goal scoring, and utilizing Poisson regression to predict the
2018-19 season outcomes. Lastly, in Section 6, we give a quick summary
of our results as well as discuss possible future work related to this
research.

\hypertarget{data}{%
\section{Data}\label{data}}

The first dataset for our investigation simply consists of match final
scores of all Premier League games from its inaugural competition, the
1992-93 season, to the last fixture of 2018-19 season. The main
attributes of this dataset are the season, the home and away teams, and
the number of goals scored by each team. We rely on
Football-Data.co.uk's data (Football-Data 2020), which contains all
Premier League match final scores from 1993 to 2019. Each season has its
own data file, and we read in and then join the individual datasets
together to get our desired data table. We utilize this data to model
the number of goals scored and then to make predictions of the 2018-19
season, using three different subset of seasons: 1) data from all
seasons prior to 2018-19, 2) data from only the 2010s, and 3) data from
all seasons, but assigning more weight to more recent competitions.

\hspace{5mm} To obtain the data for the first two season subsets, we
simply filter out the seasons that don't belong to the year ranges from
the initial table. For the assigning weight simulation method, our
weight allocation approach is very simple, as we let the weight number
be equivalent to the number of times the data for a particular season is
duplicated. We have decided that the previous five years before 2018-19
are almost all that matter. Thus, every season from 1992-93 to 2012-13
are given weight 1, then the weight increases by 1 for each one of
2013-14, 2014-15, and 2015-16. After that, we have the 2 most recent
years left and we multiply the weight by 2. Our weight values are
depicted in Figure 1.

\begin{center}\includegraphics{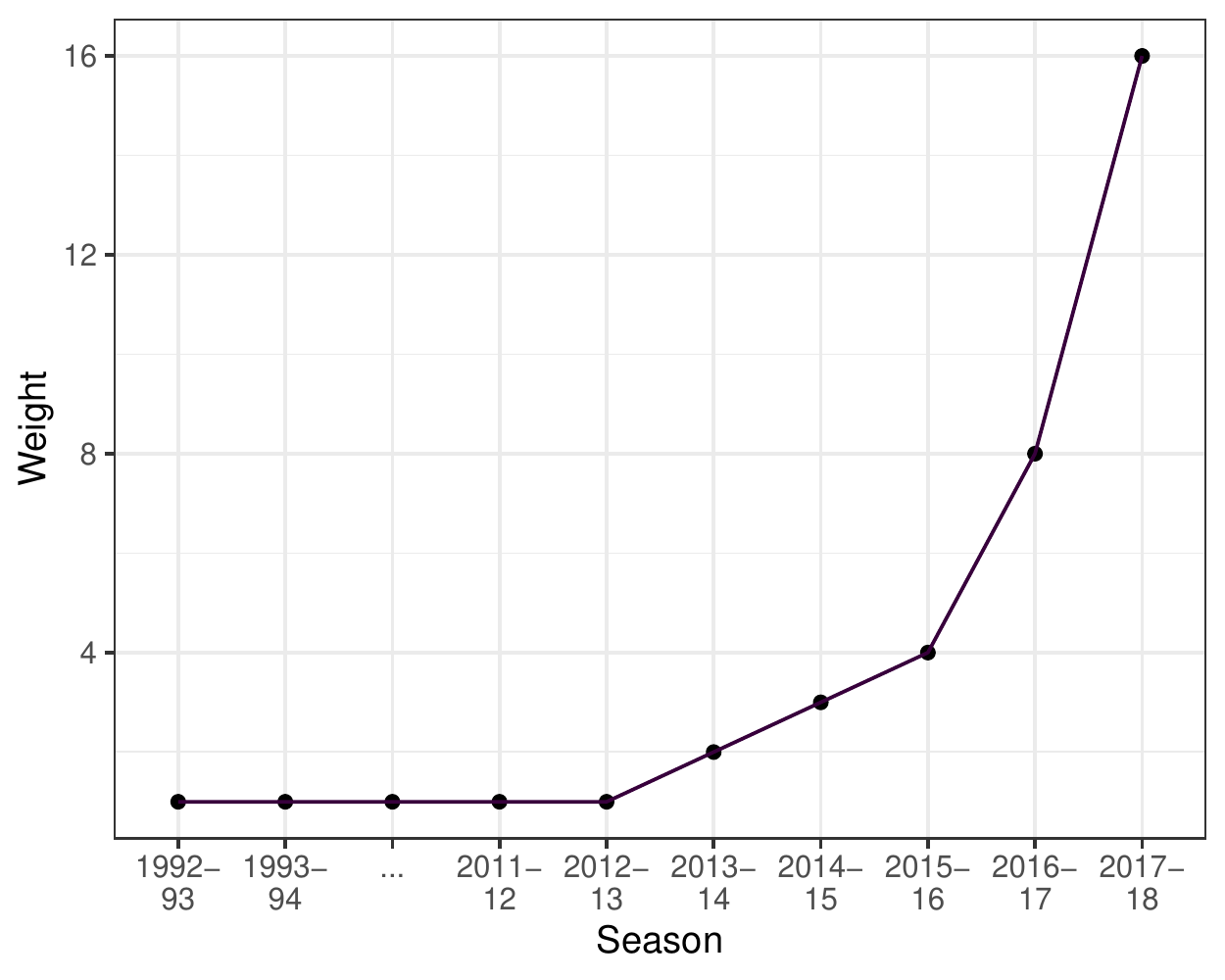} \end{center}

\vspace{-3mm}

Figure 1: Weight values across seasons from 1992-93 to 2017-18 for
predicting 2018-19 season outcomes.

\vspace{2mm}

\hspace{5mm} Finally, since we are also interested in examining goal
scoring time and the time between goals in the EPL, data on these two
topics for Manchester United, a well-known Premier League club, are
collected and stored in a spreadsheet by ourselves. This data set
contains five columns: minutes, which is the point of time during a
match at which a goal is scored; matchweek, which is the fixture number
of each game; the stoppage time in minutes for both halves of each game,
and finally, the time between goals, which is the difference in minutes
between the scoring events, taking into account stoppage time. We
collected these five variables for all Manchester United goals during
their 2018-19 Premier League campaign ({``2018--19 Manchester United
{F.C}. Season''} 2019). In addition to the initial attributes, we create
a new variable for normalizing the goal scoring minutes by dividing each
one of them by the total minutes of their respective game. The reason
for re-scaling the minutes is because the match total time varies, since
we also take into account stoppage time of football games, which means
some matches take place for a longer time than others.

\hypertarget{methodology}{%
\section{Methodology}\label{methodology}}

\hypertarget{goal-scoring-and-the-poisson-process}{%
\subsection{Goal Scoring and The Poisson
Process}\label{goal-scoring-and-the-poisson-process}}

The Poisson process (Dobrow 2016) is a stochastic process used to model
the occurrence of phenomena over a continuous interval, which in most
cases represents time. There are several characteristics of the Poisson
process that can be observed, including, the number of events happening
in a given time period; the time between those events; and when (at what
point of time) the events occur. Playing a huge role in the Poisson
process is the Poisson distribution (Asimow and Maxwell 2015), which
deals with the number of occurrences of an event in a fixed period of
time, with a rate of occurrence parameter \(\lambda\). Named for French
mathematician Siméon Denis Poisson, the Poisson distribution is a
discrete probability distribution that expresses the number of
occurrences of an event over a given period of time. The probability
density function of a Poisson random variable \(X\) with parameter
\(\lambda\) is given by

\begin{equation} 
 \displaystyle p_X(x) = \frac{{e^{ - \lambda } \lambda ^x }}{{x!}} ; \ x = 0,1,2,... \textrm{ and  } \lambda > 0,
\end{equation}

where \(X\) represents the number of occurrences of an event in a given
unit time period, and \(\lambda\) is the constant rate of occurrence per
time period. The mean and variance of our Poisson random variable \(X\),
denoted by \(\mu_X\) and \(\sigma^2_X\) respectively, are

\begin{equation} 
 \mu_X = \lambda \textrm{ and } \sigma^2_X = \lambda.
\end{equation}

\hspace{5mm} Another key distribution in this process is the exponential
distribution (Asimow and Maxwell 2015), which has a strong connection
with the Poisson distribution, in that if the number of occurrences per
interval of time are illustrated by Poisson, then the description of the
length of time between occurrences are provided by the exponential
distribution. If we have a non-negative random variable \(X\) that is
the time until the next occurrence in a Poisson process, then \(X\)
follows an exponential distribution with probability density function

\begin{equation} 
 \displaystyle f_X(x) = \lambda e^{-\lambda x} = \frac{1}{\beta} e^{-\frac{1}{\beta} x}; \ x \ge 0,
\end{equation}

where \(\lambda\) represents the average rate of occurrence and
\(\beta\) is the average time between occurrences. The mean and variance
of an exponentially distributed random variable X are

\begin{equation} 
 \displaystyle \mu_X = \frac{1}{\lambda} = \beta \textrm{ and } \displaystyle \sigma^2_X = \frac{1}{\lambda^2} = \beta^2.
\end{equation}

\hspace{5mm} Furthermore, there is a connection between Poisson and
another famous probability distribution - the continuous uniform
distribution (Asimow and Maxwell 2015). If a Poisson process contains a
finite number of events in a given time interval, then the unordered
times, or locations, or positions, or points of time at which those
events happen are uniformly distributed on that continuous interval. The
continuous uniform distribution is a probability distribution with
equally likely outcomes, meaning that its probability density is the
same at each point in an interval \([A,B]\). A continuous random
variable \(X\) is uniformly distributed on \([A,B]\) if its probability
density function is defined by

\begin{equation} 
 \displaystyle f_X(x) = \frac{1}{B-A}; \ A \le x \le B.
\end{equation}

In addition, \(X\) has mean and variance

\begin{equation} 
 \displaystyle \mu_X = \frac{A + B}{2} \textrm{ and } \displaystyle \sigma^2_X = \frac{(B-A)^2}{12}.
\end{equation}

\hspace{5mm} We postulate that goal scoring in football can be modeled
by a Poisson process. According to the characteristics described above,
if goal scoring for a club happens at a certain rate in a given time
period, then a Poisson distribution can be used to model the number of
goals scored. Additionally, the waiting time (in minutes) between
successive goals can be described using an exponential distribution.
Moreover, the time positions (or ``minute marks'') in a game at which
scoring events transpire may be uniformly distributed. We will explore
these relationships in more detail in Section 4.

\hypertarget{simulating-and-predicting-season-outcomes-using-poisson-regression}{%
\subsection{Simulating and Predicting Season Outcomes using Poisson
Regression}\label{simulating-and-predicting-season-outcomes-using-poisson-regression}}

Our second goal of this research is to use the method of Poisson
regression to predict the outcomes for EPL matches. Poisson regression
is a member of a broad class of models known as the Generalized Linear
Models (GLM) (Fox 2015). A generalized linear model has the general form

\begin{equation} 
 \displaystyle E(Y_i) = \mu_i = g^{-1}(\beta_0 + \beta_1 X_{i1} + \beta_2 X_{i2} + \ldots + \beta_k X_{ik}).
\end{equation}

\hspace{5mm} There are three main components to a generalized linear
model:

\begin{enumerate}
\def\labelenumi{\arabic{enumi}.}
\item
  A random component, indicating the conditional distribution of the
  response variable \(Y_i\) (for the \(i\)th of \(n\) independently
  sampled observations), given the values of the explanatory variables.
  \(Y_i\)'s distribution must be a member of an exponential family, such
  as Gaussian, Binomial, Poisson, or Gamma.
\item
  A linear predictor
  (\(\beta_0 + \beta_1 X_1 + \beta_2 X_2 + \ldots + \beta_k X_k\)),
  which is a linear combination of the predictors (the \(X\)'s), with
  the \(\beta\)'s as the regression coefficients to be estimated.
\item
  A canonical link function \(g(\cdot)\), which transforms the expected
  value of the response variable, \(E(Y_i) = \mu_i\), to the linear
  predictor.
\end{enumerate}

\hspace{5mm} Poisson regression models are generalized linear models
with the natural logarithm as the link function. It is used when our
response's data type is a count, which is appropriate for our case since
our count variable is the number of goals scored. The model assumes that
the observed outcome variable follows a Poisson distribution and
attempts to fit the mean parameter to a linear model of explanatory
variables. The general form of a Poisson regression model is

\begin{equation} 
 \displaystyle ln(\mu_i) = \beta_0 + \beta_1 X_{i1} + \beta_2 X_{i2} + \ldots + \beta_k X_{ik}.
\end{equation}

\hspace{5mm} To make predictions for Premier League matches and to
determine what would happen in the 2018-19 season using Poisson
regression, we fit two models to get the scoring rates for every EPL
team, 1) at home, and 2) away from home. Here we are interested in
evaluating the model equation at different values of the explanatory
variables. Since the link function for Poisson regression is the natural
log function, we would back-transform the equation with the
corresponding exponential function. This will then give us the home and
away mean (expected) scoring rates for every EPL club, aggregated across
all opponents.

\hspace{5mm} After that, we executed a large number of simulations, to
get the hypothetical 2018-19 season results and then analyzed and
compared what we got for each of the three subsets of season mentioned
in the previous section. For each subset of data, we performed 10000
simulations, and this was accomplished by randomly generating the match
final score for every team matchup, using the clubs' average scoring
rates that we got from fitting the Poisson regression models, which
returns a random integer for each team's number of goals scored. In
addition, the number of points for every match outcome based on the
teams' number of goals scored were also calculated (see Table 1), as a
side gets 3 points if they score more than their opponent, 1 point if
the final score is a tie, and 0 points if the opposing roster has more
goals. For each simulated season (out of 10000 total for each method),
we tallied up the points, calculated the goal differentials, and
obtained the final standings for EPL clubs (see Table 2). From this
information, we kept track of various metrics for EPL clubs and utilized
them to evaluate and compare the models and their predictions, which
will be discussed in the next section.

\vspace{1mm}

Table 1: Simulation table of 2018-19 EPL matches. The two columns
\texttt{HomeScore} and \texttt{AwayScore} indicate the number of goals
scored by the home and away clubs, randomly generated from their average
scoring rates. The points each team receives versus their opponent (3
for a win, 1 for a draw, 0 for a loss) are determined from the match
final score.

\begin{table}[ht]
\centering
\scalebox{0.82}{
\begin{tabular}{lrlrrrrr}
  \hline
HomeTeam & HomeRate & AwayTeam & AwayRate & HomeScore & AwayScore & HomePoints & AwayPoints \\ 
  \hline
Man City & 1.832 & Southampton & 0.989 & 1 & 0 & 3 & 0 \\ 
  Chelsea & 1.968 & Huddersfield & 0.632 & 3 & 0 & 3 & 0 \\ 
  Chelsea & 1.968 & Everton & 1.054 & 4 & 0 & 3 & 0 \\ 
  Bournemouth & 1.474 & Man United & 1.706 & 0 & 2 & 0 & 3 \\ 
  Bournemouth & 1.474 & Everton & 1.054 & 0 & 3 & 0 & 3 \\ 
   \hline
\end{tabular}
}
\end{table}

\newpage

Table 2: Final standings for a simulated 2018-19 season. The team ranks
are arranged by total points, followed by their goal differential (goal
scored minus goal conceded).

\begin{longtable}[]{@{}rlrrr@{}}
\toprule
Rank & Team & Played & Points & GD \\
\midrule
\endhead
1 & Leicester & 38 & 69 & 24 \\
2 & Arsenal & 38 & 64 & 11 \\
3 & Fulham & 38 & 64 & 7 \\
4 & Man United & 38 & 62 & 23 \\
5 & Chelsea & 38 & 60 & 13 \\
6 & Bournemouth & 38 & 59 & 11 \\
7 & Man City & 38 & 58 & 4 \\
8 & Brighton & 38 & 56 & -1 \\
9 & Everton & 38 & 52 & -2 \\
10 & Watford & 38 & 52 & -2 \\
11 & Liverpool & 38 & 51 & 4 \\
12 & West Ham & 38 & 51 & -3 \\
13 & Wolves & 38 & 51 & -10 \\
14 & Tottenham & 38 & 49 & -4 \\
15 & Cardiff & 38 & 48 & -1 \\
16 & Crystal Palace & 38 & 45 & -15 \\
17 & Newcastle & 38 & 44 & -8 \\
18 & Southampton & 38 & 40 & -15 \\
19 & Burnley & 38 & 38 & -18 \\
20 & Huddersfield & 38 & 37 & -18 \\
\bottomrule
\end{longtable}

\hypertarget{results}{%
\section{Results}\label{results}}

\hypertarget{goal-scoring-and-the-poisson-process-1}{%
\subsection{Goal Scoring and The Poisson
Process}\label{goal-scoring-and-the-poisson-process-1}}

For our first analysis of the relationship between the number of goals
scored and the Poisson distribution, we used Manchester United (MU) as
our case of inspection. Our question here was ``Does MU's distribution
of number of goals scored follow a Poisson distribution?'' Table 3 and
Figure 2 are numerical and visual summaries of Manchester United number
of goals scored in every EPL season until the final fixture of their
2018-19 campaign.

\centering

\vspace{1mm}

Table 3: Descriptive statistics of Manchester United's goals scored.

\begin{longtable}[]{@{}lrrrrrrrr@{}}
\toprule
& min & Q1 & median & Q3 & max & mean & sd & n \\
\midrule
\endhead
& 0 & 1 & 2 & 3 & 9 & 1.916185 & 1.405221 & 1038 \\
\bottomrule
\end{longtable}

\begin{center}\includegraphics{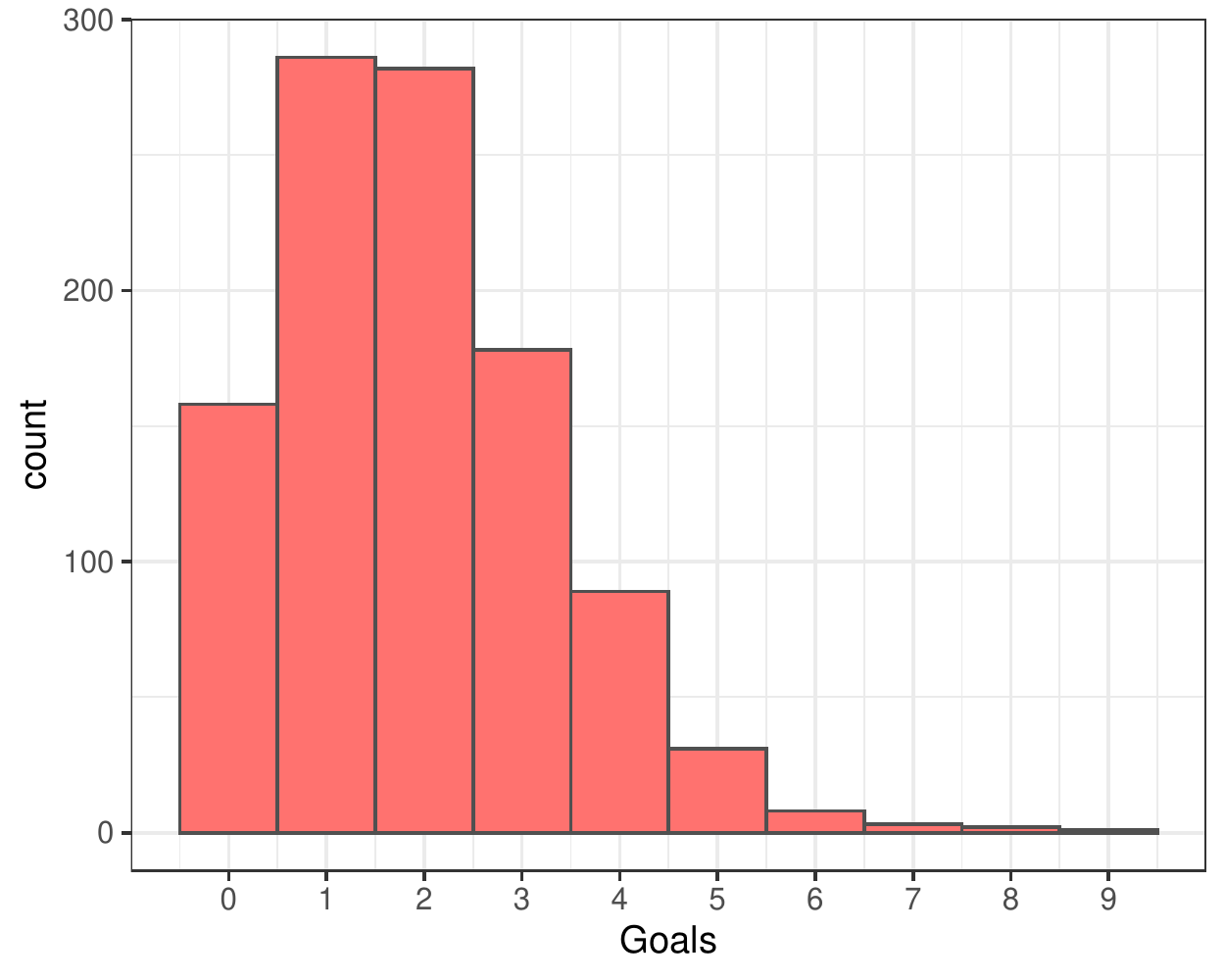} \end{center}

\vspace{-3mm}

Figure 2: Histogram of Manchester United's goals scored.

\raggedright

\vspace{2mm}

\hspace{5mm} To examine MU's number of goals scored closer, we divided
the goals variable into four levels: 0, 1, 2, 3, and 4 or more goals;
and also tally up the number of matches in real life from 1992 to 2018
having those goal values. From there, we obtained the probability of
Manchester United scoring each individual goal value, using the
probability density function of a Poisson distribution with mean
\(\lambda = 1.916\) (see Table 3); and then multiplied the probabilities
with the total number of matches of \(n = 1038\) (see Table 3), to get
the expected number of matches for each number of goals category. The
counts for the observed and expected number of matches are very close to
one another, as illustrated by Table 4 and Figure 3, which is a
significant validation for us. The results from a Chi-square
goodness-of-fit test (\(\chi^2 = 0.3805\), \(p\)-value \(= 0.984\))
further confirm that there is no significant difference between the
data's and the empirical distribution, hence the number of goals scored
by Manchester United is consistent with a Poisson distribution.

\vspace{1mm}

Table 4: Observed and expected frequencies of the number of matches for
each goal value, alongside their Poisson probabilities.

\begin{longtable}[]{@{}lrrr@{}}
\toprule
Goals & Pr\_Goals & Observed & Expected \\
\midrule
\endhead
0 & 0.147 & 158 & 153 \\
1 & 0.282 & 286 & 293 \\
2 & 0.270 & 282 & 280 \\
3 & 0.173 & 178 & 180 \\
4 or more & 0.128 & 134 & 133 \\
\bottomrule
\end{longtable}

\begin{center}\includegraphics{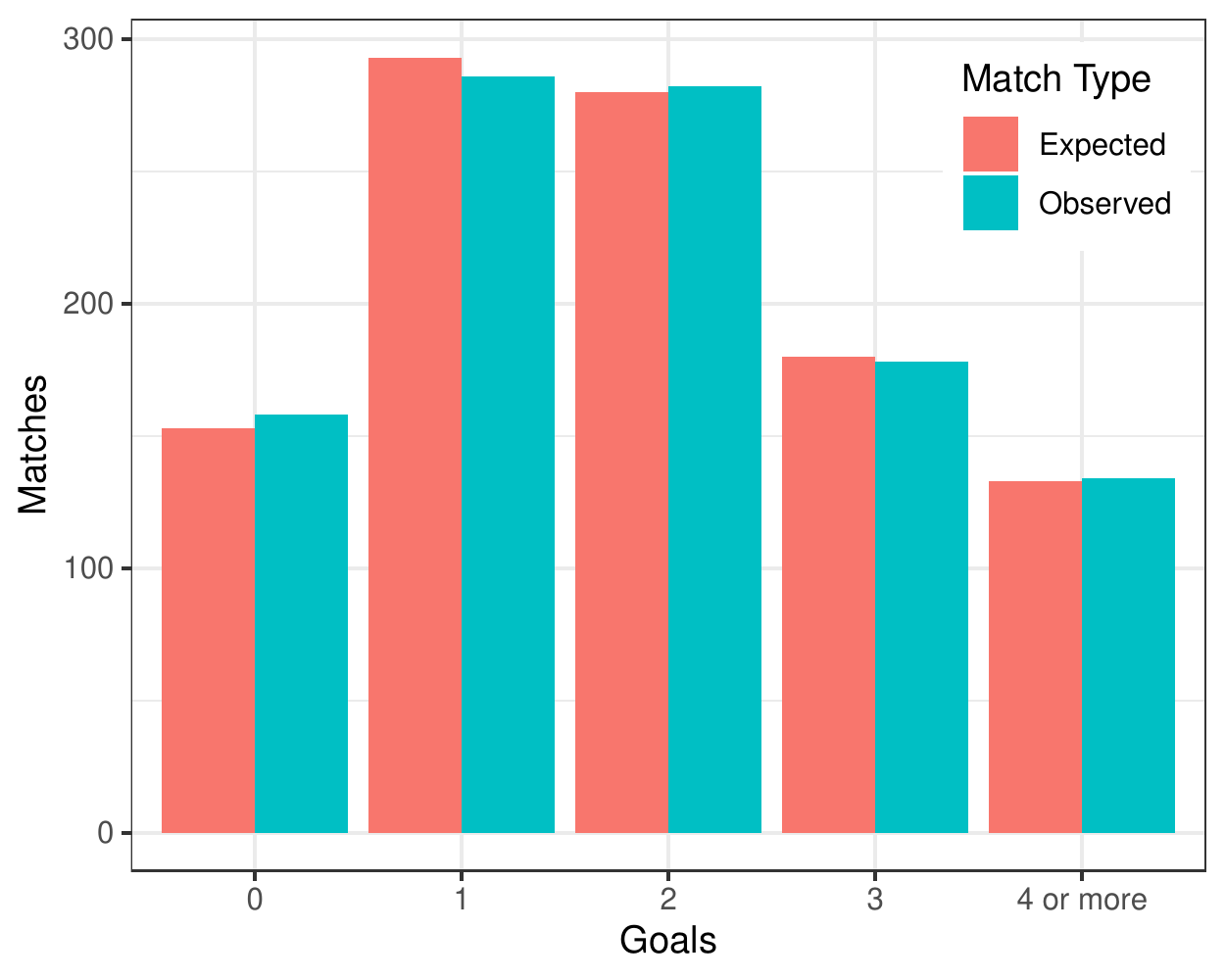} \end{center}

\centering

\vspace{-3mm}

Figure 3: Side-by-side bar graph comparing the observed and expected
matches.

\vspace{5mm}

\raggedright

\hspace{5mm} We were also interested in verifying the connections
between the time between goals in a season and the exponential
distribution, and the re-scaled goal scoring minutes in a match and the
standard uniform distribution. We continued to use Manchester United to
investigate these topics and explore their goal scoring time data
described in Section 2. We first plotted the cumulative distribution
curve of our data and compare them with the reference distributions
(exponential and uniform), as shown in Figures 4 and 5. As we can see,
the differences between the curves in each figure are marginal, which is
a validation and gives us more reason to believe in our claim that our
goal scoring data are consistent with the specified distributions. After
that, we conducted a Kolmogorov-Smirnov goodness-of-fit test to confirm
the fit between our data and the hypothetical distributions -
exponential and uniform. Based on the test statistics and \(p\)-values
(\(D = 0.0892\), \(p\)-value \(= 0.6789\) for Time between goals
vs.~Exponential and \(D = 0.0854\), \(p\)-value \(= 0.7305\) for Time
location of goals vs.~Uniform), there is insufficient evidence to
support a conclusion that our data for the time between goals and
minutes of scoring are not consistent with the exponential and uniform
distributions. Therefore, the time between goals and re-scaled scoring
time for Manchester United in the 2018-19 EPL season are exponentially
and uniformly distributed, respectively.

\begin{center}\includegraphics{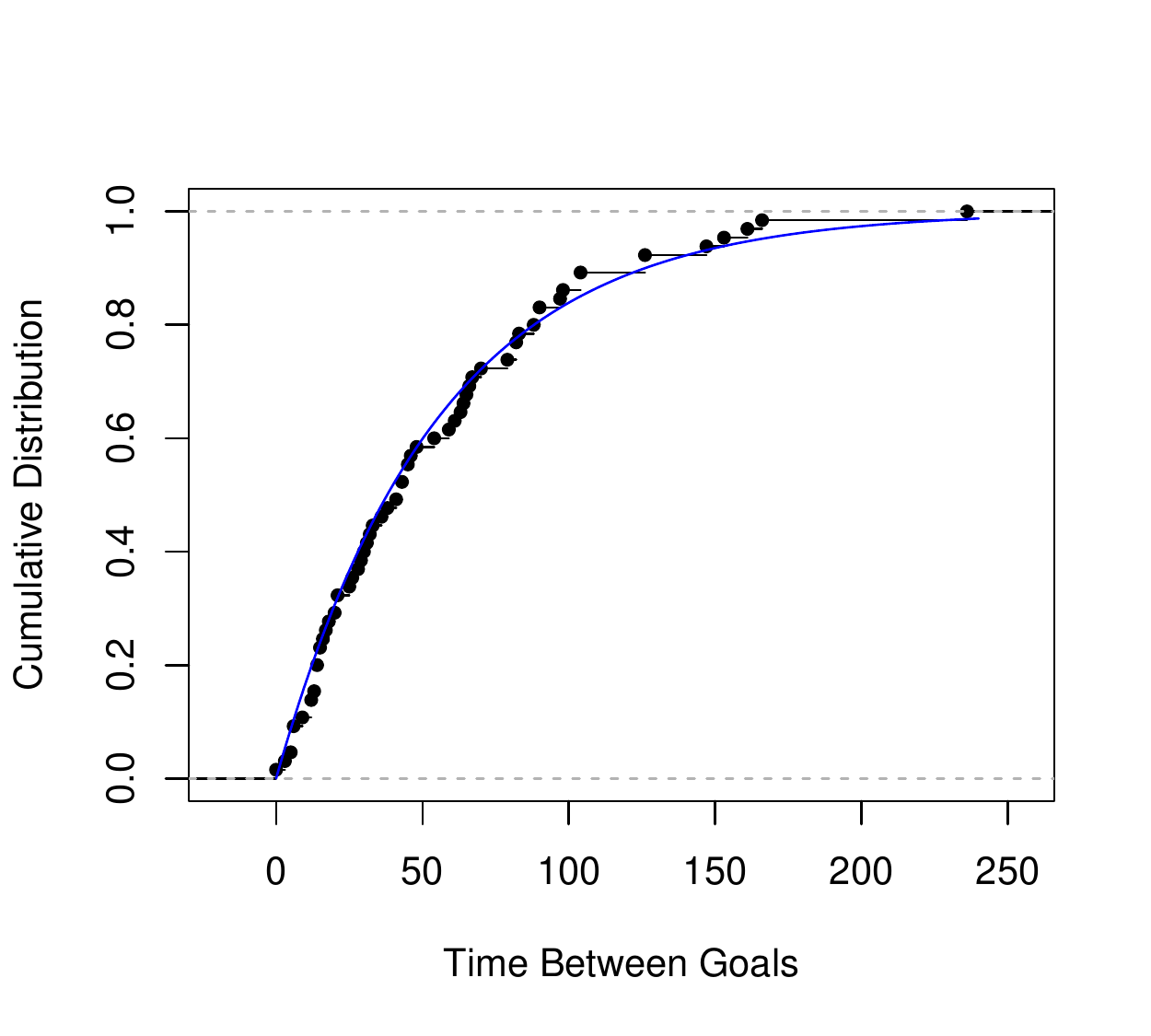} \end{center}

\vspace{-7mm}

Figure 4: Cumulative distribution curves of time between goals and the
exponential distribution.

\vspace{-15mm}

\begin{center}\includegraphics{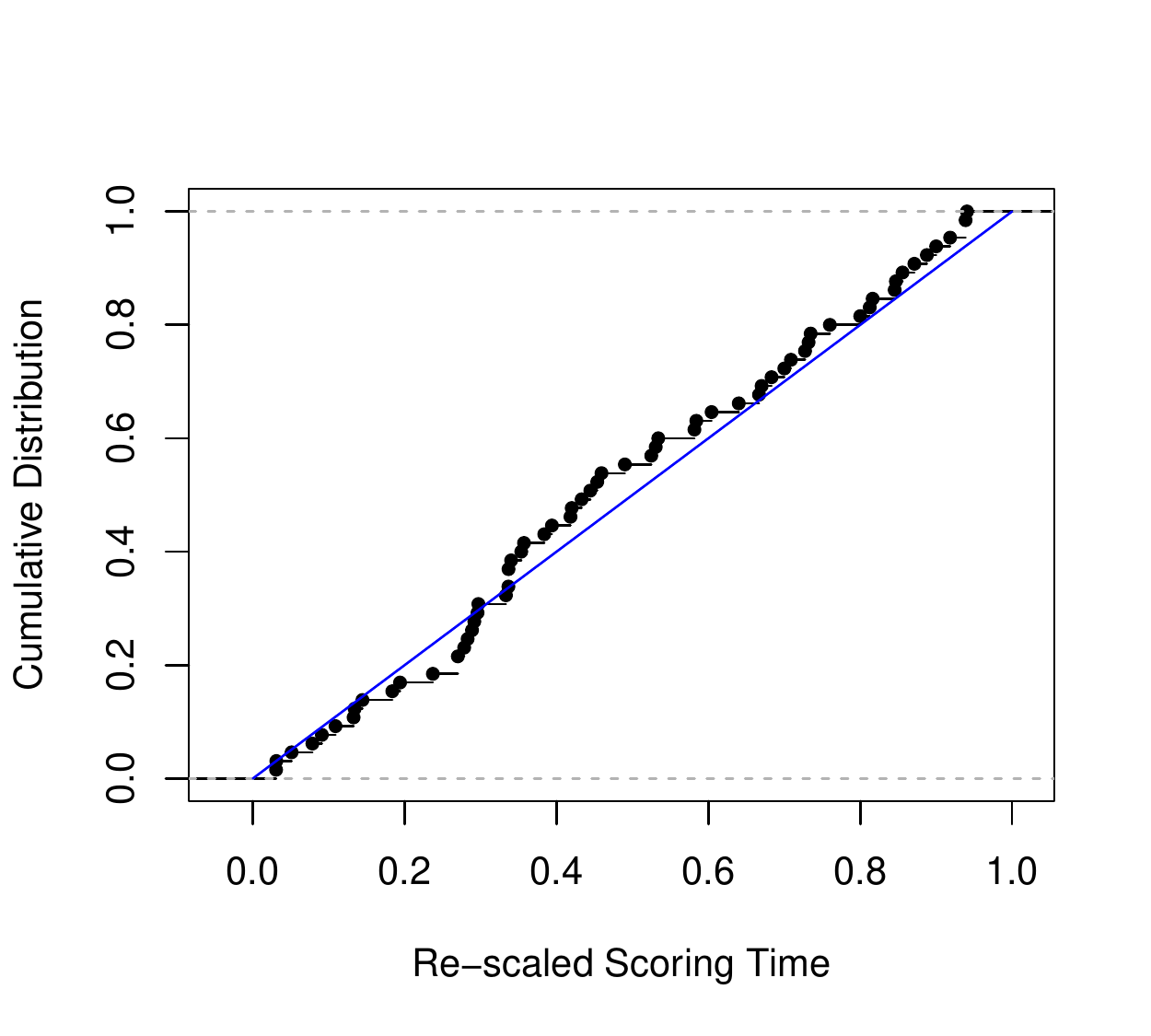} \end{center}

\vspace{-7mm}

Figure 5: Cumulative distribution curves of the re-scaled goal scoring
minutes and the standard uniform distribution.

\newpage

\hypertarget{simulating-and-predicting-season-outcomes-using-poisson-regression-1}{%
\subsection{Simulating and Predicting Season Outcomes using Poisson
Regression}\label{simulating-and-predicting-season-outcomes-using-poisson-regression-1}}

In this section, we discuss the results of our Poisson regression models
and simulations described in Section 3. After fitting the models and
conducting the simulations, there are various analyses we could do, from
comparing how teams' chances of winning differ across the three subsets
of data used for simulation, to analyzing which method gives us the most
consistent predictions compared to the actual outcomes. In this paper,
we focus on the title race and the relegation battle in the Premier
League, namely, the chances of winning the league title and getting
relegated for EPL teams.

\hspace{5mm} We first look at the chances of finishing first in the EPL
table at the end of the 2018-19 season for the ``Big 6'' in English
football, which includes Manchester United, Liverpool, Arsenal, Chelsea,
Manchester City, and Tottenham (see Table 5). Overall, it is clear that
MU's chance of having higher ranks decreases drastically if we
prioritize recent data over just using data from all seasons. This is
consistent with the club's situation in real life, as MU has not been
doing well lately, but was winning a lot prior to the 2010s. Their
in-town rivals, Manchester City, on the other hand, have much higher
percentages of winning the league in 2018-19 if we focus on data from
recent years. In fact, they were the champions of the 2018-19 EPL
season. Similar to Manchester United, our results for Manchester City
also make sense, due to the fact that City has emerged into an EPL title
contender over the last decade, but they were just an average team
dating back to the old 1990s and 2000s days. It is also notable that
both assigning weight to more recent years and using only data from the
2010s lower Chelsea and Arsenal's likelihood of finishing first, whereas
these two methods increase Tottenham's chance of winning the league,
though their winning chance is still the smallest in each category.

\centering

\vspace{1mm}

Table 5: Chances of winning the 2018-19 Premier League title for the Big
6.

\raggedright

\begin{longtable}[]{@{}lrrr@{}}
\toprule
Team & All Seasons & 2010s & Assign Weight \\
\midrule
\endhead
Arsenal & 19.68 & 15.05 & 14.07 \\
Chelsea & 14.28 & 11.70 & 9.03 \\
Liverpool & 12.50 & 10.96 & 17.61 \\
Man City & 7.00 & 41.71 & 38.09 \\
Man United & 36.99 & 10.47 & 10.53 \\
Tottenham & 3.91 & 7.38 & 8.34 \\
\bottomrule
\end{longtable}

\hspace{5mm} Next, we investigate the likelihood of getting relegated
after the 2018-19 season for EPL clubs. The relegation zone, or the last
three places in the final rankings, is where no teams in the Premier
League want to end up at the end of the season, because after each
season, the bottom three clubs get relegated to the second highest
division of English football. The results from our three subsets of
season data (see Table 6) are pretty consistent with each other, with
Huddersfield and Cardiff essentially being the two ``locks'' to play in
the English football division below the EPL in the following season.
Brighton and Burnley also have high chances of being in the bottom
three, although in reality, both of these teams successfully remained in
the league for another year. The third team that got dismissed in 2019,
Fulham, does not have high chances of relegation in any of the three
simulation methods. Unsurprisingly, Big 6 teams have the smallest
chances of getting related from the EPL at the completion point of the
2018-19 season.

\centering

\vspace{1mm}

Table 6: Chances of getting relegated after the 2018-19 season for
Premier League teams.

\raggedright

\begin{longtable}[]{@{}lrrr@{}}
\toprule
Team & All Seasons & 2010s & Assign Weight \\
\midrule
\endhead
Huddersfield & 69.15 & 72.91 & 71.03 \\
Cardiff & 49.99 & 54.38 & 53.47 \\
Brighton & 41.65 & 45.03 & 44.03 \\
Burnley & 31.60 & 43.38 & 33.34 \\
Watford & 26.28 & 17.24 & 15.82 \\
Wolves & 22.04 & 14.22 & 22.32 \\
Crystal Palace & 17.54 & 14.42 & 10.65 \\
Fulham & 9.87 & 7.91 & 11.74 \\
West Ham & 7.46 & 6.73 & 6.39 \\
Southampton & 6.49 & 5.23 & 11.65 \\
Leicester & 5.53 & 1.75 & 2.39 \\
Bournemouth & 4.76 & 5.46 & 5.80 \\
Everton & 3.48 & 2.17 & 3.63 \\
Newcastle & 2.40 & 8.78 & 7.41 \\
Tottenham & 1.02 & 0.16 & 0.08 \\
Man City & 0.38 & NA & NA \\
Chelsea & 0.15 & 0.09 & 0.13 \\
Arsenal & 0.10 & 0.04 & 0.06 \\
Liverpool & 0.10 & 0.05 & 0.02 \\
Man United & 0.01 & 0.05 & 0.04 \\
\bottomrule
\end{longtable}

\hspace{5mm} On a related note, the 40-point safety rule (Davis 2018) is
an interesting myth associated with the EPL's relegation zone. Since the
league's reduction in the number of teams to twenty clubs before the
start of the 1995-96 season, there have been only three times that a
squad got relegated despite hitting the 40-point mark. They are West Ham
in 2002-03 with 42 points, and Bolton and Sunderland both with 40 points
at the end of 1997-98 and 1996-97 seasons respectively. This mythical
40-point mark has been crucial for the relegation battle for many years,
as subpar teams often view getting there as their ``security blanket''
for remaining in the top division of English football. From Table 7,
using data from 2010s and assigning weight to recent years give us
significantly less number of both teams and seasons that violate the
40-point rule than using data from every season prior to 2018-19.
Therefore, the 40-point safety rule seems to hold much better for the
two simulations that focus on recent data than the one with data from
all previous seasons. This actually makes sense, since in the past two
decades in reality, teams with 40 or more points at the end of EPL
seasons all survived from relegation, as the last time this rule did not
happen was the 1997-98 season.

\vspace{1mm}

Table 7: 40-point safety rules comparison between the three subsets of
season data. We tally up the total number of teams as well as distinct
simulated seasons with teams being relegated while having at least 40
points for each simulation method.

\begin{longtable}[]{@{}lrr@{}}
\toprule
Subset & Seasons & Teams \\
\midrule
\endhead
All Seasons & 3434 & 4424 \\
2010s & 2066 & 2470 \\
Assign Weight & 2346 & 2889 \\
\bottomrule
\end{longtable}

\hypertarget{conclusion-and-discussion}{%
\section{Conclusion and Discussion}\label{conclusion-and-discussion}}

Overall, we have found that Premier League goal scoring fits the
characteristics of a Poisson process. Our first result was that a
Poisson distribution can be used to predict the number of matches with
each number of goals scored. Additionally, the time between each
individual goal in a season can be described by an exponential
distribution. We also have evidence that the normalized goal scoring
time positions after are uniformly distributed. We also used different
sets of data prior to the 2018-19 Premier League season, namely, data
from all seasons before, data from only the 2010s, and data from all
previous years but assigning more weight to recent competitions, to
predict what would happen in the 2018-19 season. We got each team's goal
scoring rate at home and away from home by doing Poisson regression, and
then performed simulations using those rate parameters. Different team
metrics like how many points each team got and what place each team
finished were being kept track of from the simulations, and then we make
use of those variables to analyze and compare our models of different
season data subsets.

\hspace{5mm} In the future, there are additional topics we could
explore, including:

\begin{enumerate}
\def\labelenumi{\arabic{enumi}.}
\item
  Besides the number of goals scored, there are many other factors that
  can be used to determine outcomes of football matches. In future
  research, we could use various factors to predict goal scoring and
  find out if they will be as helpful as using just number of goals. We
  could look into variables that are likely to contribute to the
  outcomes of Premier League football matches such as clean sheets,
  possession time, pass accuracy, shots on target, and numerous other
  soccer statistics. On top of that, we could compare different models
  with different predictors and evaluate them to find out which set of
  variables best predicts league outcomes, and then use them to simulate
  and predict match results.
\item
  In football and many other sports, team performance tends to vary
  throughout a season and across seasons. Some Premier League teams have
  the tendency of getting hot in early months, some clubs reach their
  peak during the middle period of the season, and a few others are more
  likely to do better at the season's home stretch. Winning and losing
  streaks are also important factors in sports, as some clubs are
  streaky, while others tend to be more consistent. Thus, in future
  research, we could apply match results of EPL teams from past games
  within the season, and maybe find a way to emphasize winning and
  losing streaks, to predict the outcome of later matches. As a follow
  up, we could investigate a model's performance throughout the season.
  Some models may work better and predict more accurate results at
  certain times in the year than others.
\item
  In addition to predicting match results, another popular application
  of statistical modeling in sports analytics is determining betting
  odds. We could use the probabilities from our Poisson regression
  models and simulations to calculate the odds of possible game outcomes
  for different team matchups. We could also look into and compare
  different types of bets such as over and under, money line wager, or
  point spread; determining if it is a good idea to bet on a match, and
  if so, how much profit we could win.
\end{enumerate}

\hypertarget{acknowledgement}{%
\section*{Acknowledgement}\label{acknowledgement}}
\addcontentsline{toc}{section}{Acknowledgement}

This work was completed as the author's senior honors thesis, in partial
fulfillment of the requirements for earning Departmental Honors in
Mathematics at Wittenberg University in Springfield, Ohio. The author
would like to express his special gratitude and thanks to his advisor,
Professor Douglas M. Andrews, for his many ideas, suggestions and
guidance throughout the research process. The author would also like to
thank the Department of Mathematics and Computer Science at Wittenberg
University for providing him the valuable knowledge during his
undergraduate career, and for giving him the opportunity to participate
in the Departmental Honors Program.

\hypertarget{supplementary-material}{%
\section*{Supplementary Material}\label{supplementary-material}}
\addcontentsline{toc}{section}{Supplementary Material}

All of the materials related to this research are publicly available on
GitHub at \url{https://github.com/qntkhvn/eplgoals}.

\hypertarget{references}{%
\section*{References}\label{references}}
\addcontentsline{toc}{section}{References}

\hypertarget{refs}{}
\begin{CSLReferences}{1}{0}
\leavevmode\vadjust pre{\hypertarget{ref-mu1819}{}}%
{``2018--19 Manchester United {F.C}. Season.''} 2019.
\url{https://en.wikipedia.org/wiki/2018-19_Manchester_United_F.C._season}.

\leavevmode\vadjust pre{\hypertarget{ref-probstat}{}}%
Asimow, L. A., and M. M. Maxwell. 2015. \emph{Probability and Statistics
with Applications: A Problem Solving Text}. 2nd ed. ACTEX Publications.

\leavevmode\vadjust pre{\hypertarget{ref-epl40pt}{}}%
Davis, M. 2018. {``Do Premier League Teams Really Need 40 Points to
Avoid Relegation?''} \url{https://www.bbc.com/sport/football/43049564}.

\leavevmode\vadjust pre{\hypertarget{ref-stochastic}{}}%
Dobrow, R. P. 2016. \emph{Introduction to Stochastic Processes with
{R}}. John Wiley \& Sons, Inc.
\url{https://doi.org/10.1002/9781118740712}.

\leavevmode\vadjust pre{\hypertarget{ref-footballdata}{}}%
Football-Data. 2020. {``Footbal-Data.co.uk.''}
\url{https://www.football-data.co.uk/englandm.php}.

\leavevmode\vadjust pre{\hypertarget{ref-fox2015applied}{}}%
Fox, J. 2015. \emph{Applied Regression Analysis and Generalized Linear
Models}. SAGE Publications.

\leavevmode\vadjust pre{\hypertarget{ref-footyhist}{}}%
Johnson, B. 2015. {``Association Football or Soccer.''}
\url{https://www.historic-uk.com/CultureUK/Association-Football-or-Soccer}.

\leavevmode\vadjust pre{\hypertarget{ref-eplhist}{}}%
{``Premier League Explained.''} 2020.
\url{https://www.premierleague.com/premier-league-explained}.

\end{CSLReferences}

\end{document}